# IoTSim-Edge: A Simulation Framework for Modeling the Behaviour of IoT and Edge Computing Environments

Devki Nandan Jha*, Khaled Alwasel*, Areeb Alshoshan*, Xianghua Huang*, Ranesh Kumar Naha†,
Sudheer Kumar Battula†, Saurabh Garg†, Deepak Puthal*, Philip James*, Albert Zomaya‡,
Schahram Dustdar§, Rajiv Ranjan*
*Newcastle University, United Kingdom.
† University of Tasmania, Australia.
‡The University of Sydney, Australia.
§TU Wien, Austria.

**Abstract**— With the proliferation of Internet of Things (IoT) and edge computing paradigms, billions of IoT devices are being networked to support data-driven and real-time decision making across numerous application domains including smart homes, smart transport, and smart buildings. These ubiquitously distributed IoT devices send the raw data to their respective edge device (e.g. IoT gateways) or the cloud directly. The wide spectrum of possible application use cases make the design and networking of IoT and edge computing layers a very tedious process due to the: (i) complexity and heterogeneity of end-point networks (e.g. wifi, 4G, Bluetooth); (ii) heterogeneity of edge and IoT hardware resources and software stack; (iv) mobility of IoT devices; and (iii) the complex interplay between the IoT and edge layers. Unlike cloud computing, where researchers and developers seeking to test capacity planning, resource selection, network configuration, computation placement and security management strategies had access to public cloud infrastructure (e.g. Amazon and Azure), establishing an IoT and edge computing testbed which offers a high degree of verisimilitude is not only complex, costly and resource intensive but also time-intensive. Moreover, testing in real IoT and edge computing environments is not feasible due to the high cost and diverse domain knowledge required in order to reason about their diversity, scalability and usability. To support performance testing and validation of IoT and edge computing configurations and algorithms at scale, simulation frameworks should be developed. Hence, this paper proposes a novel simulator IoTSim-Edge, which captures the behaviour of heterogeneous IoT and edge computing infrastructure and allows users to test their infrastructure and framework in an easy and configurable manner. IoTSim-Edge extends the capability of CloudSim to incorporate the different features of edge and IoT devices. The effectiveness of IoTSim-Edge is described using three test cases. The results show the varying capability of IoTSim-Edge in terms of application composition, battery-oriented modeling, heterogeneous protocols modeling and mobility modeling along with the resources provisioning for IoT applications.

———————————— ◆ ————————————

## 1 INTRODUCTION

Advances in IoT have a transformative impact on society and the environment through a multitude of application areas including smart homes, smart agriculture, manufacturing and healthcare. To achieve this, an ever increasing number of heterogeneous IoT devices are continuously being networked to support real-time monitoring and actuation across different domains. Cisco predicts that 50 billion IoT devices are going to be connected by 2020. Traditionally, the enormous amount of data, generally known as the big data, is sent to the cloud by IoT devices for further processing and analysis. However, the centralized processing in cloud is not suitable for numerous IoT applications due to the following reasons: (i) Some applications require close coupling between request and response [1], (ii) Delay incurred by the centralized cloud-based deployment is unacceptable for many latency-sensitive applications [2], (iii) There is a higher chance of network failure and data loss, and (iv) Sending all the data to cloud may drain the battery of the IoT device at a faster rate [3]. This has led to the evolution of edge computing solution [4]. Though some of the existing, literature distinguishes between edge and fog computing, following [5] in this paper we abstract both to these, relatively new, paradigms of computing to be part of edge layer.

The introduction of edge computing addresses these issues by providing the computational capacity in a near proximity to the data generating devices. Smart edge devices such as Smart phone, Raspberry Pi, UDOO board, etc. supports local processing and storage of data on a widespread but smaller scale. However, the constituent devices in edge computing are heterogeneous as each one may have specific architecture and follows particular protocols for communication. Unlike the cloud where the location of a datacenter (data centre in the UK) is fixed, edge and IoT devices can be mobile and change location frequently. In addition to this, edge and IoT devices are powered by batteries, solar (or a combination of the two) or continuously connected to an external power supply as compared to cloud datacenter which is always connected to a stable power source. To exploit the advantages offered by edge computing, it is necessary to understand the features and capabilities of edge and IoT devices along with their composition in a proposed IoT data analysis framework. The diversity of underlying IoT and edge devices, data types and formats, communication mediums, application scope, functional complexity, and programming models makes the evaluation very challenging and time consuming.

Evaluating the framework in a real environment gives the best performance behaviour, however, it is not always possible as most of the test frameworks are in development. Even if the infrastructure is available, it is very complex to perform the experiment as setting it up requires knowledge of all the associated IoT and edge devices which is not intuitive. Running multiple experiments to test a desired framework requires reconfiguration of multiple devices and changes to required parameters quickly becomes untenable due to the volume of changes needed. Additionally, performing the experiment in the real environment is very expensive due to the set up and maintenance cost incurred. Since, the real environment is dynamic, it is very hard to reproduce the same result for different iterations which may lead to mis-interpretation of the evaluation. All these challenges hinders the use of real environment for benchmarking the edge computing environments. To overcome this issue, another feasible alternative is the use of *simulators*. The simulators offer a window of opportunity for evaluating the proposed hypothesis (frameworks and policies) in a simple, controlled and repeatable environment. A simulation environment must mimic the key complexity and heterogeneity of real networks and support multiple scenarios that affect real IoT deployments.

### 1.1 Challenges

Simulating and modeling a realistic IoT scenario is very challenging due to various reasons [6] such as (i) Variety of IoT devices need to be combined with edge device and cloud to satisfy the requirements of an application; (ii) Modeling networking graph between diverse type of IoT and edge computing device in an abstract manner can be very challenging; (iii) Modeling data and control flow dependencies between IoT and edge layers to support diverse data analysis work-flow structure is non-trivial; (iv) capacity planning across edge computing layer is challenging as it depends on various configuration parameters including data volume, data velocity, upstream/downstream network bandwidth, to name a few; (v) The communication between IoT and edge device is very different from cloud datacenter communication, which are generally based on wired and/or wireless protocol. The connectivity between IoT and edge computing layers, as we discuss later in the paper, can be diverse. Hence, they are very difficult to model in an abstract way while not loosing the expressiveness, i.e. lower level details related to protocol latency, impact of protocols on battery discharge rate of underlying IoT device, etc; (vi) Mobility is another important parameter of IoT devices as sensors embedded to many physical devices are moving. Since the range of edge device is limited, the movement of sensor may leads to handoff. Also, the data sent to an edge device for processing may not be in the current range of IoT device. Thus for receiving the processed data, an edge to edge communication is required. Modeling the mobility and handoff for large number of IoT devices with varying velocity is very challenging; (vii) Dynamicity of IoT environment leads to addition and removal of IoT and edge devices very frequently. This may be caused by numerous factors e.g. device failure, network link failure. Modeling the scalability of IoT devices with heterogeneous features at a fast rate is very challenging; and (viii) Since IoT environment is an emerging area, new applications might be developed in future. It is very important for a simulator to allow users to customize and extend the framework based on their requirements. Making a general simulator that allows easy customization is very challenging.

Different simulators are proposed in the literature. Simulators such as CloudSim [7] and GreenCloud [8] are specific for cloud environment, however, EdgeCloudSim [9] and iFogSim [10] are proposed for edge computing environment. However, from the best of our knowledge, we could not find any simulator that addresses all the above challenges.

### 1.2 Contributions

This paper aims to build a novel simulator, IoTSim-Edge that allow users to evaluate the edge computing scenario in a easily configurable and customizable environment. IoTSim-Edge is build on the existing simulators which tries to capture the complete behaviour of IoT and edge computing infrastructure development and deployment. Especially it covers all the above discussed challenges in a seamless manner. Additionally, the proposed simulator can be easily used to analyze various existing or futuristic IoT applications. In particular, the proposed simulator is able to model the following scenarios:

- New IoT application graph modeling abstraction that allows practitioners to define the data analytic operations and their mapping to different parts of the infrastructure (e.g. IoT and edge) (see §4.1).
- Abstraction that supports modeling of heterogeneous IoT protocols along with their energy consumption profile. It allows practitioners to define the configuration of edge and IoT devices along with the specific protocols they support for networking. Details are presented in §4.2.
- Abstraction that supports modeling of mobile IoT devices (see §4.3). It also captures the effect of handoff caused by the movement of IoT devices. To maintain a consistent communication, IoTSim-Edge supports a cooperative edge-to-edge communication that allows the transfer the processed data of the respective IoT device by one edge via another edge.

Outline This paper is organized as follows. Section 2 presents a brief background of IoT infrastructure environment and the architecture of edge computing. An illustration of the architecture of our proposed simulator, IoTSim-Edge is given in Section 3. It also explains the implementation details of IoTSim-Edge. Three case studies evaluating IoTSim-Edge is presented in Section 4 while the recent related work comparing IoTSim-Edge with the existing simulator is given in Section 5. Section 6 concludes the paper giving some future work suggestions.

## 2 IoT and Edge Computing

This section discusses the background information of the IoT environment in terms of the modeling challenges. It also gives the general architecture of edge computing considered for modeling by the proposed simulator.

### 2.1 IoT Environment

IoT can be defined as the "sophisticated sensors and chips which are embedded in the physical things, objects and living beings surrounding us and connected through the Internet to monitor and control the connected things" [11]. Numerous IoT applications improve our daily lifecycle in different vertical of domains such as smart homes, smart health care, Industry 4.0 and disaster management. IoT functionality is delivered by six main elements namely sensing, identification, communication, computation, services and semantics [12]. IoT devices sense the environment while scattered ubiquitously capturing the physical and environmental information. IoT devices are identified based on the application requirements and techniques employed to implement the applications. The computation process is distributed across IoT device, edge and cloud datacenter based on the desired functional and Quality of Service (QoS) parameters of the application. To achieve this data is sent from IoT device to edge and further from edge to cloud using different communication protocols. The computational result can be used to make some decisive operation to achieve the desired application process.

Consider a simple example of smart home that controls all the devices of a home and eases the life of the inhabitant. The IoT devices are sensors embedded to all the devices such as refrigerator, heater/cooler, light bulb and car while the edge devices are gateways and mobile phones. The smart home system uses private cloud datacenter resources. Home devices are connected to the gateway using light weight protocol, Bluetooth Low Energy (BLE) using CoAP application protocol for data transmission. Mobile phone is connected to 4G while gateways are connected to WIFI for data transmission to private cloud. When the resident person leave for office, the smart home system automatically switch off the light bulbs and heater/cooler. The system also checks the refrigerator for available egg or milk and sends a message to the person to bring those things. Based on the location information got from the person's phone and car, the system restarts the bulbs and adjusts the room temperature.

Modeling such a realistic IoT application requires a combination of sensors, actuators and edge devices on a large scale with different operating environments. It is a complicated task due to the heterogeneous characteristics of IoT and edge devices and requires continuous optimization for resource provisioning, allocation, migration and fault tolerance during the application processing. Again the implementation is specific for only one application. Giving a generic model for IoT application that can be modeled for any IoT application requires a level of abstraction along with specific details for that application. The main challenges are in terms of modeling application composition, network protocols, mobility of IoT devices and energy consumption in the form of battery drainage as discussed below.

**Application Composition** An IoT application consists of a sequence of operations performed on sensed data. It can be represented in variety of ways, however, this paper follows [13] to represent an IoT application as a directed acyclic graph (DAG) of MicroElements (MELs). Each MEL is an abstract component of the application that represents the resources, services and data altogether in the form of microservice, microdata, microcomputing or microactuation [13]. Modeling an application as a DAG of MEL is very challenging as we need to encapsulate multiple components together. Also, the sequence of MEL is very important as it represents the data and control flow at the abstract level.

**Communication protocols** In The IoT environment, network connectivity and messaging protocols play a significant role in the communication. By the inherent characteristic of the IoT environment it has a complex network interaction between different IoT environment components. Table 1 shows the common communication protocols employed by the IoT environment. Based on the distance, type of devices and specified constraints, any of the given protocol can be utilized by the IoT and edge devices for data transfer. Mobile devices need to leverage different protocols as compared to static ones. Modeling these protocols within the application graph is very challenging.

Additionally, various messaging protocols are available to facilitate the data transfer from sensors to edge devices and further to cloud servers. Table 2 briefs few commonly used protocols for this purpose. Transferring data using any of these protocols affects the system performance in significantly different ways. A single messaging protocol will not be able to satisfy the requirements of complex IoT use case. Hence, it is required to share different protocols for different devices and at different layers. Modeling these scenarios with handshaking between various protocols is very challenging

**Mobility of IoT devices** IoT devices embedded with cars or smart phones supports mobility to assist users in more flexible ways. Considering the range of edge device as fixed, mobile IoT device can move from the range of one edge device to another causing handoff. The handoff may be hard or soft depending on the speed of IoT device and the signal range of edge device [14]. In order to simulate the mobility in a realistic way, we need a number of features e.g. IoT device speed and acceleration/deceleration, motion path, edge range intersection, topological maps [15]. Incorporating all these features for the realistic mobility simulation is a very complex task because of a large number of data points with highly dependent characteristics and relationships. Moreover, the data transfer may fail at any time when the IoT device is moving from one location to another location due to the weak strength of the signal.

**Battery drainage** Most of the IoT devices are powered by battery which is limited and need to be recharged at particular period of time. It is very important for these devices to sustain the battery for longer duration especially for application where it is not easy to recharge e.g. sensor in river or in a disaster place (earthquake, landslip, etc.). Sustaining battery for longer time saves a lot of cost and which is vital for every application. Sending data at different rate and using different communication protocols cause the battery drainage at different rates. Therefore, it is very important to monitor the battery consumption in different case scenarios. Simulating the battery drainage of IoT device is very complex task because of the non-linear effect while drainage of battery and the fact that each device hardware supports different network connectivity and communication protocols. Hence, providing multi-level abstraction for energy characterization with fine accuracy and tight energy consumption bounds for different devices based on the various factors such as hardware, sensing requests, and communication protocols is a complex task.

TABLE 1: Overview of different communication protocols (L : Low, M : Medium, H : High, VS : Very Short, VL: Very Low).

| Protocol | Data Rate | Distance | Power efficiency | Reliablity | Cost | Service |
|---|---|---|---|---|---|---|
| Bluetooth Low energy (BLE) | Approx. 0.27 Mbps (M) | Approx 100m (M ) | L | H | L | Wearable devices Smart connected devices |
| WIFI | Approx. 54Mbps (H) | Approx 50m (M) | M | M | L | Home IoT Office IoT Smart cities |
| 4G LTE H | Approx 12 Mbps (L) | Large | H | H | H | Agriculture Industries Transportation fleets |
| Zigbee | Approx. 250kbps (L) | Approx. 100m ( M) | L | H | L | Indoor Asset tracking Home Automation |
| Long Range (LoRa) | Approx. 50kbs ( L) | Approx. miles (L) | L | H | M | Smart city Energy Management Supply chain Management |
| Sig Fox | Approx. 1kbos (VL) | Approx. several miles (L) | VL | H | M | Environmental Sensors Smart meters |
| NFC | Approx. 42kbps (M) | Approx. 20cm (VS) | VL | H | VL | Contactless Payment Transaction. Local Asset Tracking |

TABLE 2: Overview of different messaging protocols.

| App. layer Protocol | Restful | Trans. layer Protocol | QoS | Architecture | Security | Header Size (Bytes) | Sync. |
|---|---|---|---|---|---|---|---|
| CoAP | Yes | UDP | Yes | Pub-Sub, Req-Res | DTLS | 4 (min) | async/sync |
| XMPP | No | TCP | No | Pub-Sub, Req-Res | SSL | - | async |
| MQTT | No | TCP | Yes | Pub-Sub | SSL | 2 | async/sync |
| AMQP | No | TCP | Yes | Pub-Sub | SSL | 8 | async |
| HTTP | Yes | TCP | No | Req-Res | SSL | - | async |

## 2.2 Architecture of IoT-Edge Computing

Figure 1 shows the architecture of IoT-Edge computing. The IoT infrastructure consists of mainly 2 components Sensing nodes and Actuator nodes. Sensing nodes will collect the information of surroundings through sensors and send the information for processing and storage. Actuators will be activated based on the analysis of the data. The communication layer is responsible for data transfer to/from IoT devices, edge devices and cloud. Different communication protocols are available for data transfer as shown in Table 1 and 2.

The next layer is for edge infrastructure which consists of different type of edge devices such as Arduino, Raspberry Pi. These devices can be accessed transparently with the help of different type of virtualization and containerization mechanisms. It provides the infrastructure for the deployment of the raw data generated by the sensing nodes. In many cases when the edge is enable enough to process the data, it does not need to send the data to cloud for further processing. Finally, the result is sent back to the actuator for performing the particular action.

The application or services layer consists of different services that can be directly accessible to the users. These applications will be accessed via a subscription model. The example services are a smart home, smart-City, smart healthcare and smart transportation.

The application management layer will manage the applications deployed in the edge environment. This layer is responsible for the application composition where an application is decomposed into a DAG of MELs which abstracts both software and data [16]. Each MEL can be deployed across the edge or cloud datacenter. It also manages the QoS requirements along with load balancing and fault tolerance handling. It offers other management services such as resource management, storage management and device management. Thus services provided by this layer will ensure the user's QoS requirements satisfaction.

Existing application deployment and scaling techniques developed for other distributed computing environments such as cloud or grid are not suitable for the new IoT-Edge environment. This is because of the diverse characteristics of smart devices along with the mobility feature and the modern application architecture that have a strict dependency and requires distributed processing. Depending on the application type, a variety of collaboration between IoT, edge and cloud is required to achieve the desired QoS requirements. Hence, the development of new application deployment and scaling techniques are required. It is necessary to test and validate these techniques before actual deployment. However, testing these new techniques in the real environment with different conditions is very time consuming and expensive. Moreover, due to the distributed ownership of the devices, testing requires multiple access mechanisms which makes it even more complicated. Therefore, a simulation framework such as IoTSim-Edge simulator, which supports the deployment of an application which evaluates the performance of different techniques, scenarios under different conditions is required. Moreover, evaluating the techniques under different scenarios and conditions can be done with minimal cost in the simulation environment.

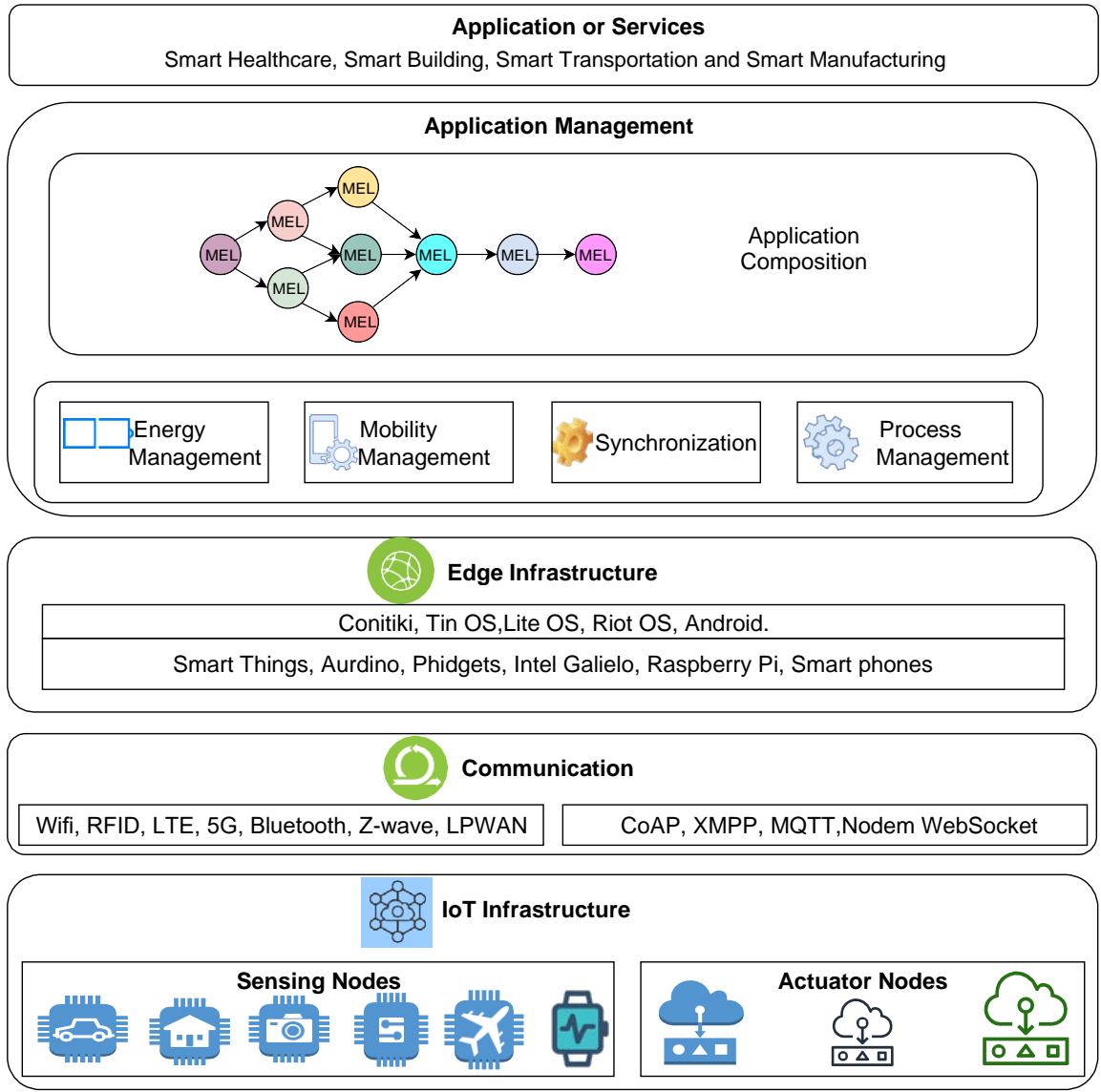

Fig. 1: The Architecture of IoT-Edge Computing.

## 3 IoTSim-Edge Architecture

The architecture of the proposed simulator consists of multiple layers as shown in Figure 2. A brief description of each component is presented in this section.

IoTSim-Edge is built on the top of CloudSim [7] simulation tool. CloudSim provides the underlying mechanisms for handling communication among subscribed components (e.g. broker, edge datacenter, IoT resources) using an event management system. The core components of CloudSim are extended to represent the edge infrastructure in line with edge's features and characteristics. IoT resources layer contains different types of IoT devices (e.g. car sensor, motion sensor) where each one has their own features and behaviours along with performing different operations of sensing and actuation. Sensors in the IoT device seamlessly generates data while actuators are responsible for generating the response. Traditionally, cloud resources are used for processing IoT data. But in Edge-IoT approach, sensor data is processed in the edge datacenter for faster processing time. Edge datacenter consists of heterogeneous processing devices such as smartphone, laptop, Raspberry Pi and single server machine. Edge-IoT management layer coordinates processing by receiving user request from users' layer and process the requests using the Edge-IoT resources. Resource broker facilitates the deployment process. User can provide users input through a graphical user interface (GUI) by mentioning different device configuration and policies. Edge-IoT management layer consists of several components such as Edgelet, policies, mobility, battery, synchronism, QoS, network protocols, communication protocols, transport protocols, and security protocols.

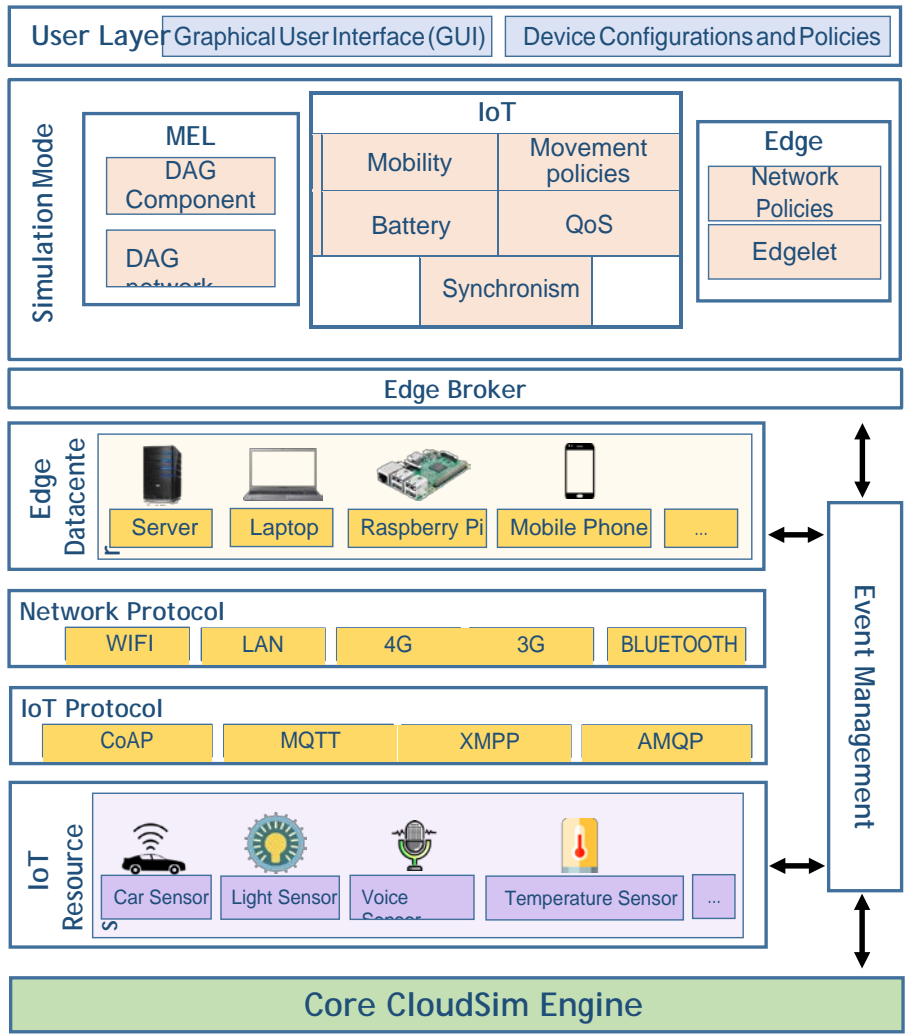

Fig. 2: The architecture of IoTSim-Edge simulator.

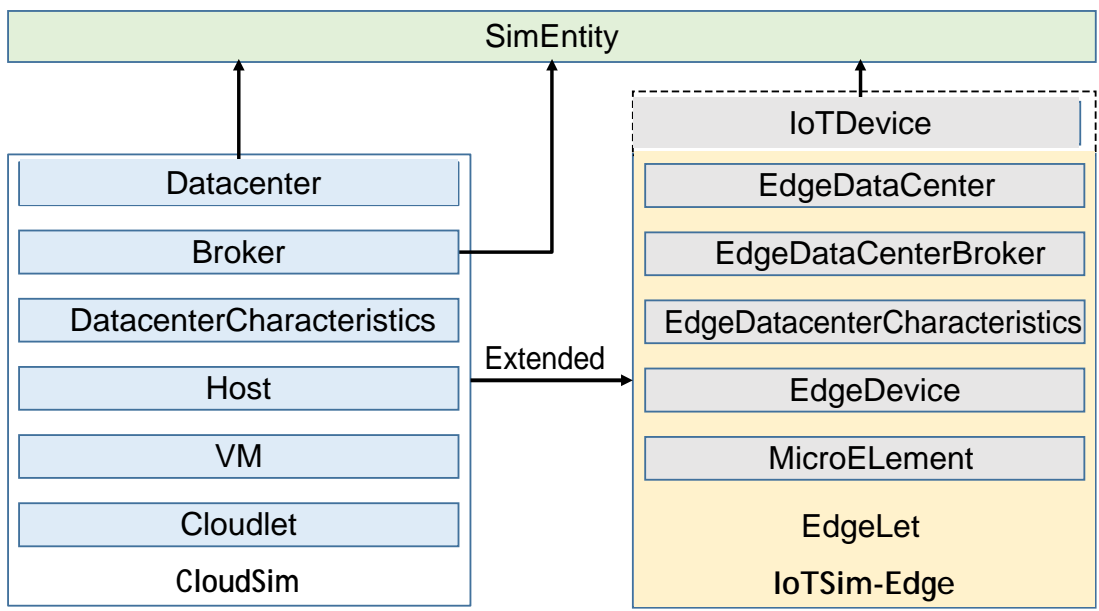

Fig. 3: Used classes of ClousSim in IoTSim-Edge plus classes use event management system

## 3.1 Design and Implementation

For the implementation of IoTSim-Edge, we extended the existing classes of CloudSim as shown in Figure 3 as well as defined numerous new classes in order to model realistic IoT and edge environments (see Figure 3 and Figure 6). Any entity that extends SimEntity class can seamlessly send and receive events to other entities when required through the event management engine. Figure 4 shows a simplified behaviour of the simulator's life-cycle as steps and actions.

For modeling an edge infrastructure, we designed and implemented numerous new classes. The main classes are EdgeDataCenter, EdgeBroker, EdgeDatacenterCharacteristics, EdgeDevice, MicroElement, and EdgeLet. The Edge-DataCenter class is responsible for establishing connection between edge and IoT devices based on the given IoT protocol (e.g. CoAP) along with performing edge resource provisioning, scheduling policies, and monitoring edge processing. Furthermore, it is also responsible for EdgeDevice creation, EdgeLet submission, setting up a network connection between edge and IoT infrastructure, checking network availability, among others. The Characteristics of EdgeDataCenter (e.g. types of policies, types of IoT protocols, types of IoT devices) is fed by EdgeDatacenterCharacteristic class. The EdgeBroker class acts on behalf of users in terms of establishing connection with edge and IoT devices, negotiating with resources, submitting IoT and edge requests, and receiving results. The EdgeDevice class presents the model of edge devices in terms of receiving and processing IoT generated data where the data processing is carried out in accordance with the given EdgeLet policy. The MicroElement (MEL) class models the abstract operation performed on IoT data on either edge or Cloud datacenters. For the current implementation of MEL, we only consider it running on edge devices. The EdgeLet class models tasks that need to be executed inside MEL. An edge device may contains a battery (e.g. mobile phone) and be moving around; therefore, the Battery and Mobility classes are designed to enable the edge device to obtain such characteristics. The Moving-Policy Class dictates the moving conditions and behaviours of edge devices.

Similarly for modeling the IoT infrastructure, we developed many new classes as shown in figure 6. The IoTdevice class mimics the behaviour of real IoT devices in terms of sensing, processing, mobility, data rate, etc. Since IoT devices are often self-powered and moving, the Battery and Mobility classes exist to empower IoT devices with such characteristics, similar to edge devices as mentioned before. The movement policy of IoT devices is directed by the MovingPolicy class; it can be extended with new moving policies according to users' requirements (e.g. velocity and location of cars' sensors). The NetworkProtocol class models well-known communication protocols (e.g. WiFi, 4G LTE) in terms of speed rate; for example, WiFi can transfer network packets at a speed of 200 Mbps while 4G LTE  at a speed of 150 Mbps. The speed rate of EdgeLet, in other words the delay time to send EdgeLet to an edge datacenter, is obtained from the NetworkPolicy class. The IoTProtocol class models the features of IoT protocol with regard  to QoS and battery consumption. As every IoT protocol (e.g. CoAP and XMPP) has different processing techniques, each one is modeled in a way that every one has different power consumption rate. More detail description of each class is given as follows.

EdgeDataCenter class This class controls the core edge infrastructure. It intercepts all incoming events and perform different operations based on the payload of the event, such as resource provisioning and submitting EdgeLet requests to respective MEL(s). It obtains and monitors the capacity of edge devices and MELs along with capturing and reporting MEL processing status to the edge broker. The class is also modeled to support a location-aware mechanism for IoT and edge devices, which helps in establishing and terminating the connection between edge and IoT devices based on the range criteria. This class also supports a power-aware technique to track the battery lifetime of edge devices. Once the battery of an edge device is fully discharged, EdgeDataCenter will automatically detach the edge device and forward unserved requests to another available edge device.

EdgeBroker class This class is a users' proxy, in which it generates users' requests with accordance to their prescribed requirements. It has a range of duties to perform, such as submitting edge and MEL provisioning requests to edge datacenter, requesting IoT devices to generate and send data to their respective edge devices, and receiving final processing results from MEL. This class supports a power-aware model for IoT devices. As EdgeBroker continuously tracks the battery consumption of IoT devices, it will disconnect the drained IoT devices from its available IoT device list.

EdgeDevice class This class behaves similar to a real edge device. It hosts several MELs and facilitates the procedure of CPU sharing mechanism via a given CPU sharing policy (e.g. time-shared, space-shared). It is also connected to a specific number of IoT devices which sends their data for processing. It can easily be attached with a battery and power draining policy based on the case when it battery-driven.

IoTDevice class This class models the core characteristics of IoT devices, particularly those that share and have in common such as generating and sending data. As every device has its own specifics (e.g. protocols, data generation rate, power consumption rate), the class is therefore extended to include the missing features of a receptive IoT device. In the current version of the simulator, there are four classes that extends the IoTDevice class: VoiceSensor, LightSensor, TemperatureSensor, and CarSensor as shown in Figure 6. Any new required type of IoT device can easily extends the IoTDevice Class and implement the new features. In the real IoT environment, every IoT device is equipped with IoT protocol (e.g. CoAP) and network protocol (e.g. 4G LTE); therefore, this class contains similar characteristics by using IoTProtocol and NetworkProtocol classes, as discussed later.

MEL class This class represents one component of IoT application graph. It represents the main processing requirement of the application component. Based on the application requirement, setEdgeOperation method can be configured. The dependency between various components can be represented using upLink and downLink which can be easily set to represent any complex application. It can be easily configured to represents any complex IoT application

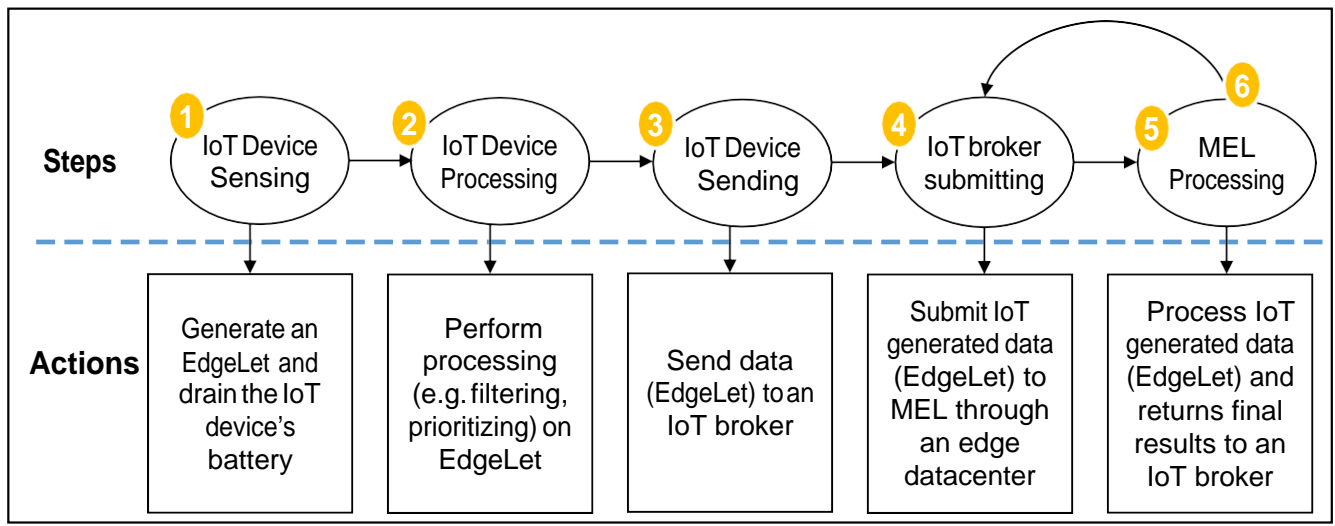

Fig. 4: An overview of IoTSim-Edge steps and actions

based on the user's requirement.

EdgeLet class This class models the IoT generated data and MEL processing data. Once IoT device establishes connection with its respective edge device(s), it generates IoT sensed data as required in a form of EdgeLet, which contains payload. The payload encapsulates required information to guide MEL on the processing stage, such as size of data to be processed, IoT device ID, and destination MEL ID. By using EdgeLet data size, the network delay required to send EdgeLet to its destination MEL can be computed taking into account the network transmission rate.

Mobility class This class contains the mobility model of IoT and edge devices. The mobility plays a vital role on IoT-Edge ecosystem where maintaining a real-time connectivity status is of importance to properly evaluate the performance. The attributes of mobility model consists of range, velocity, location, and time interval. Each attribute, excluding time interval, has two separate values to represent the horizontal and vertical direction of IoT and edge devices. Figure 7 shows an example of how such attributes are used by an edge datacenter to track the location of IoT and edge devices. Algorithm 1 shows the pseudo-code of a simple tracking technique implemented in the EdgeDataCenter class. Users can easily extend this class to implement their own mobility model.

Battery class This class models the battery characteristics of a portable IoT or edge device. The class captures the battery's behaviours and consistently computes its lifetime. In a real IoT-Edge environment, it is very difficult to know the exact power consumption for every executed task in addition to the power consumed by the device's internal components. Therefore, this class simplifies the power consumption model as an inverse proportion relationship between draining rate and battery capacity as discussed in detail in §3.2.

IoTProtocol class This class models well-known IoT application protocols. Every protocol has its own attributes and features such as power consumption rate. For our proposed simulator, the implemented classes considers four best candidates in IoT and edge computing: MQTT(Message Queue Telemetry Transport), AMQP(Advanced Message Queuing Protocol), XMPP(Extensible Messaging and Presence Protocol), and CoAP(Constrained Application Protocol) [17], however, it can be easily extended for any other considered protocol. QoS parameters are also associated with the IoT protocols which controls the acknowledgement and response method for each protocol.

NetworkProtocol class This class presents the modeling of network protocols (e.g. WiFi, 4G LTE). Implementing such models in the IoTSim-Edge framework is required to properly evaluate the performance of IoT-Edge applications. Each network type is designated with its relative network speed (e.g. 200 Mbps for WiFi, 150 Mbps for 4G LTE). By modeling the transmission rate, transmission time can obtained taken in to account the EdgeLet size.

Policies classes These classes models the policies for three activities namely, device movement, battery consumption, and network transmission. The device movement policy instruct edge datacenters to keep tracking of the movement and location of IoT and edge devices (see algorithm 1). The battery consumption policy tracks and computes the remaining power capacity of IoT and edge devices. The network transmission policy computes the time taken to transfer data from IoT device to edge device. Such policies can be extended to derive new IoT-Edge designs and solutions. All these classes can be extended to implement and test different user policies.

UserInterface class UserInterface classes provides the necessary methods to easily configure and test the IoT application development without knowing the details of the simulator. It allows a user to define all the parameters using the interactive interface which is converted to the desired configuration file. A snapshot of the user interface is shown in Figure 8.

## 3.2 Calculation and Event Processing

As shown in the sequence diagram (see Figure 9), the simulation process takes place immediately after initialising

```
"ioTDeviceEntities": [
        {
                "mobilityEntity": {
                        "movable": false,
                        "location": {
                                "x": 0.0,
                                "y": 0.0,
                                "z": 0.0
                        }
                },
                "assignmentId": 1,
                "ioTClassName": "org.edge.core.iot.TemperatureSensor",
                "iotType": "environmental",
                "name": "temperature",
                "data_frequency": 1.0,
                "dataGenerationTime": 1.0,
                "complexityOfDataPackage": 1,
                "networkModelEntity": {
                        "networkType": "wifi",
                        "communicationProtocol": "xmpp"
                },
                "max_battery_capacity": 100.0,
                "battery_drainage_rate": 1.0,
                "processingAbility": 1.0,
                "numberofEntity": 5
        }]
```

Fig. 5: An example of IoT definitions in a JSON format

---

**Algorithm 1:** Update and Track Location

1 [1] t: time, int: interval, $v_x$ : velocity in X direction, $v_y$ : velocity in Y direction, $rangeX_E$ : X range of edge device E, $rangeY_E$ : Y range of edge device E, $locationX_I$ : X coordinate of the IoT device I, $locationY_I$ : Y coordinate of the IoT device I

2 $isOut Range \leftarrow false$

3 $t_{new} \leftarrow t_{old} + int$

4 $locationX_I^{new} \leftarrow locationX_I^{old} + v_x \times int$

5 $locationY_I^{new} \leftarrow locationY_I^{old} + v_y \times int$

6 **if** $locationX_I^{new} > rangeX_E$ " $locationY_I^{new} > rangeY_E$ **then**

7 $isOut Range \leftarrow true$ **else**

8 $sendInternalEvent(I, int, updateLocation)$ **return** $isOut Range$

---

the required IoT-Edge infrastructure which is derived from a given configuration file. Figure 5 shows the test configuration file for IoT device. Once the required IoT devices and MELs (residing in edge hosts) are created, the edge broker will ask edge datacenters to establish connections between IoT devices and their respective MELs. As a result, every edge datacenter will establish requested connections and update the edge broker with the connection status for every request. As well, every edge datacenter will consistently track the location of IoT and edge devices in addition to the power consumption of every edge device during the whole lifetime of the simulation process. Note that the battery consumption of every IoT devices is maintained by the edge broker.

Every time the edge broker receives a new connection establishment (ACK), it will notify the IoT device about the connection. IoT device will then start sensing and generating data in a form of EdgeLets, transfer the EdgeLets to destination MELs, and drain its battery power according to a defined drainage rate. The IoT devices will keep repeating these actions until they reach their simulation ending time. Every time a MEL receives IoT generated data (EdgeLet), it will process the EdgeLet and returns final output to the edge broker.

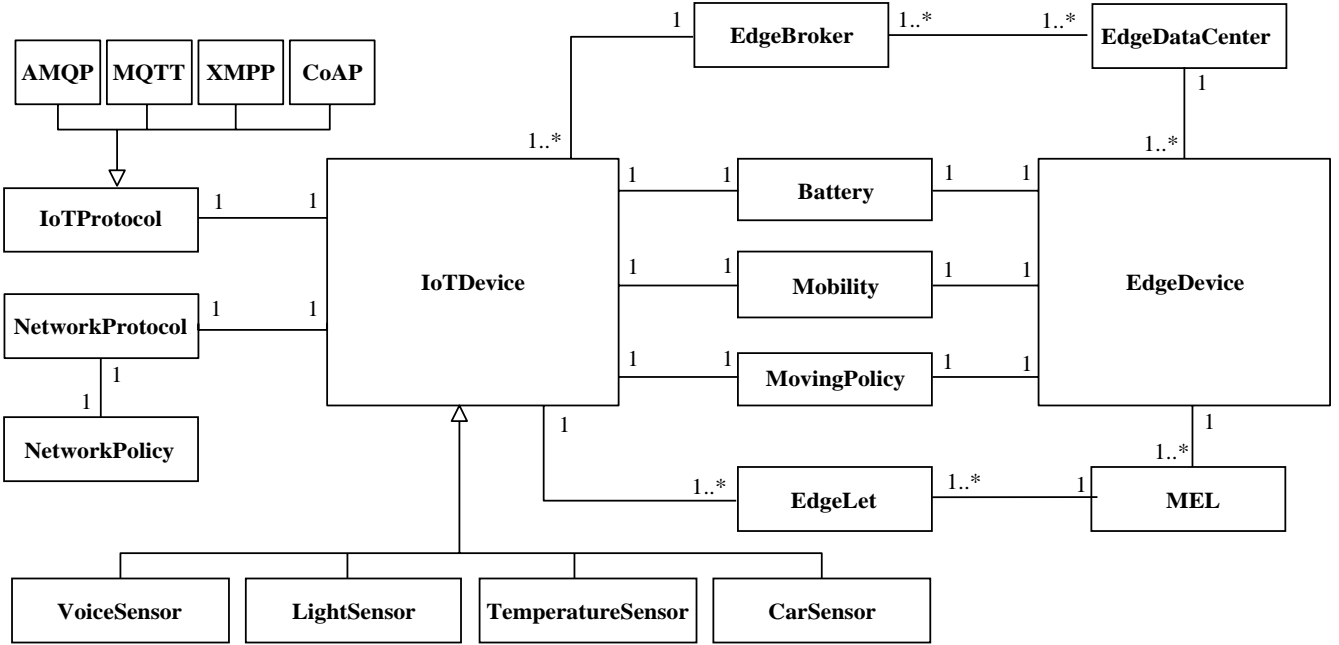

Fig. 6: Class diagram of IoTSim-Edge simulator.

TABLE 3: Configuration file for Case 1.

| IoT device | | MEL | | Edge device 1 | |
|---|---|---|---|---|---|
| Location | 0, 0, 0 | id | 1 | Type | Raspberry Pi |
| IoT type | healthcare | MIPS | 10000 | Location | 100, 0, 0 |
| Movable | false | RAM | 10000 MB | Movable | false |
| Data frequency | 1 | BW | 10000 MBPS | Signal range | 100 m |
| Data generation time | 1 | Shrinking factor | 'variable' | Max IoT device capacity | 10000 |
| Network protocol | bluetooth | Network protocol | bluetooth | Max battery capacity | 20000 units |
| IoT Protocol | COAP | Uplink | - | Battery drainage rate for processing | 0.1 units/hr |
| Max battery capacity | 300 units | Downlink | 2 | Battery drainage rate for transfer | 0.6 units/hr |

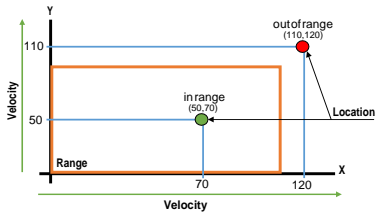

Fig. 7: How an edge datacenter tracks the location of IoT and edge devices

TABLE 4: Battery running hours using COAP and XMPP protocol

| Protocol | Battery Hours |
|---|---|
| COAP | 149.73 |
| XMPP | 119.32 |

The total processing time of a MEL depends on the data size to be processed. In our implementation, the processing is done in two step as shown below. Given the data size $DS$, first the data is converted to the number of million instructions (MI) based on the type of functionality performed as shown in Equation 1. Again based on the shrinking factor, new EdgeLet is constructed with shrinked data size which is transferred to other edge device. Along with this, a processing is performed on the data that takes $Time_{proc}$ time as shown in Equation 2. Here, MIPS is the CPU capacity in million instructions per second. The total processing time, $Total\_Proc\_Time_{MEL}$ is given by the maximum time required to either shrink or process the data as given in Equation 3.

$$MI = f(DS) \qquad (1)$$

$$Time_{proc} = \frac{MI}{CPUCapacity(MIPS)} \qquad (2)$$

$$Total\_Proc\_Time_{MEL} = max(Time_{shrink}, Time_{proc}) \qquad (3)$$

As a more detailed explanation, the updateBattery method will be called during the sensing process which updates the battery. Battery consumption directly depends on the processing data size and the underlying transmission protocol. The new battery level, $Bat_{new}$ is calculated as given in Equation 4.

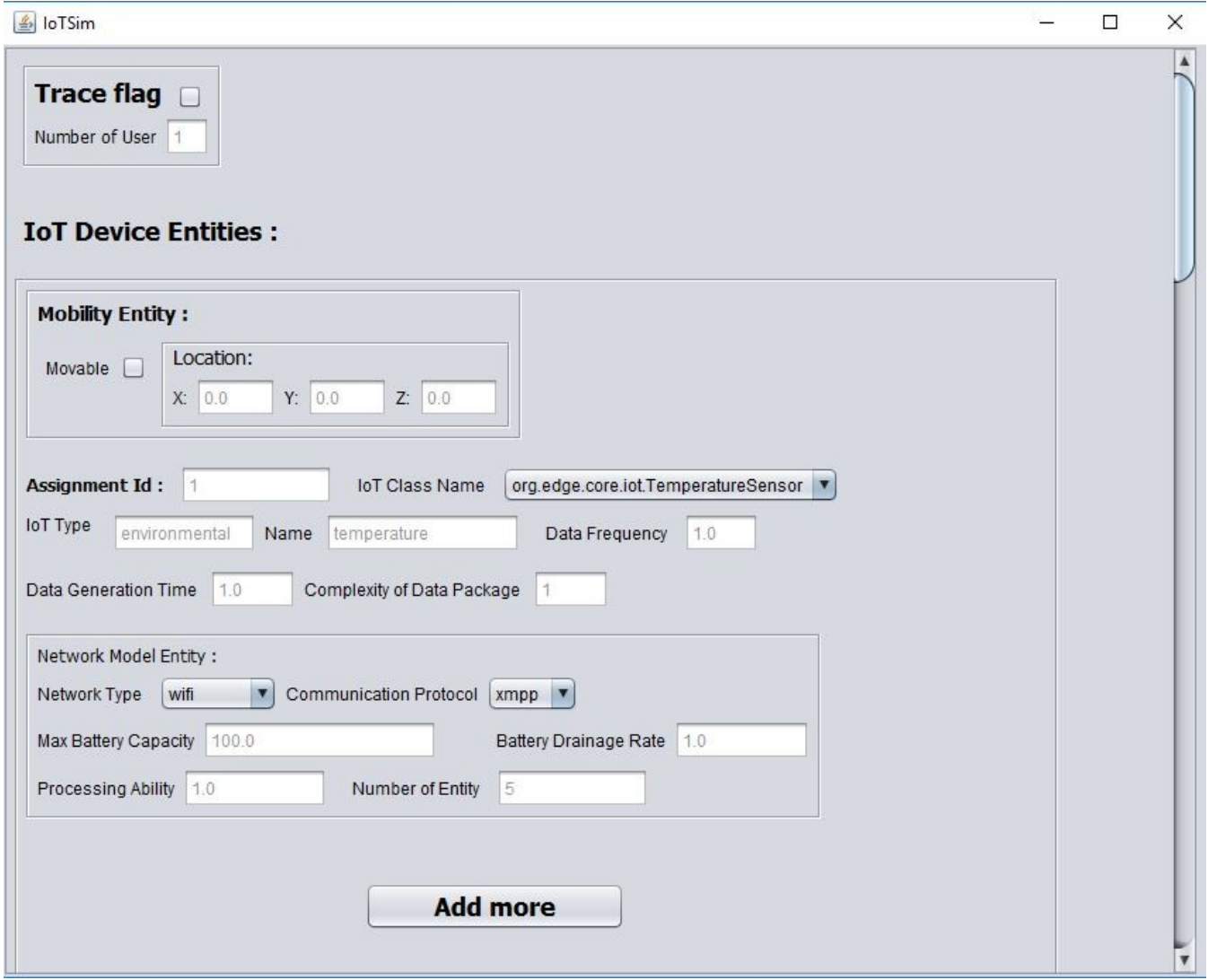

Fig. 8: Snapshot of User Interface for IoTSim-Edge

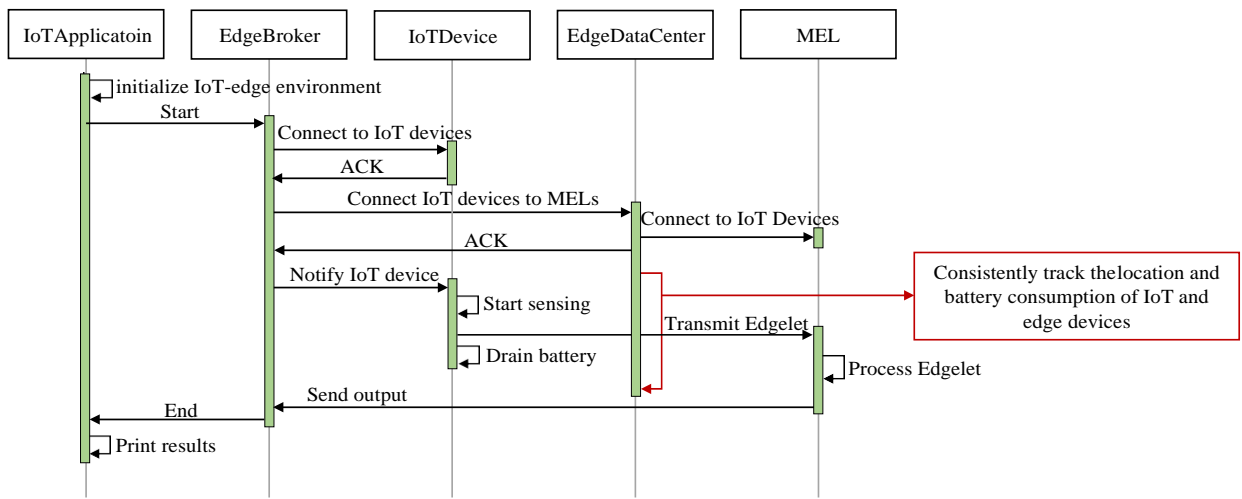

Fig. 9: Workflow of IoTSim-Edge simulator.

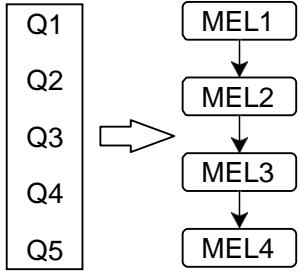

Fig. 10: A simple MEL graph for distributed deployment for healthcare query operation

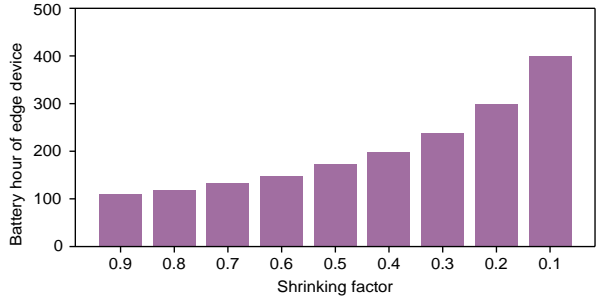

Fig. 11: Simulation result for Case 1 showing the variation of battery hours for different scenarios

$$Bat_{new} = Bat_{old} - Battery\_Consumption_{total} \quad (4)$$

$Battery\_Consumption_{total}$ is the battery consumption which is calculated as given in Equation 5.

$$Battery\_Consumption_{total} = DS[(1 - \rho) \times \lambda_{proc} + \times \rho \times \lambda_{comm}] \quad (5)$$

where, $DS$ is total data size, $\rho$ is the shrinking factor, $\lambda_{proc}$ is the drainage rate for processing and $\lambda_{comm}$ is the drainage rate for data transfer using specific protocol. Our assumption is that if the shrinking factor $\rho$, $1 - \rho$ times $DS$ is processed and $\rho$ times $DS$ is sent. The remaining battery is calculated after every processing or transfer.

If the battery of IoT device is drained, the device will be shut down. If updateBattery method finds some IoT device with available battery, the data generation event will be invoked that generates data at the defined data rate (frequency).

The generated data is sent to edge device using some specific protocol. The network transmission time depends on the protocol used as every protocol has specific data packet size and rate rate specified. The transmission time, $Time_{trans}$ is calculated as given in Equation 6.

$$Time_{trans} = \frac{N_{packet}}{data\_rate_P} \quad (6)$$

where, $N_{packet}$ is the total number of packet to be transmitted and $data\_rate_P$ is the data rate for protocol $P$. Number of packets, $N_{packet}$ depends on the max packet size allowed by the particular protocol.

Every time the data generated by the IoT device, it is immediately uploaded to the edge. The next step is checking edge device availability by the broker. If the specified edge device is not available, the broker will find a new edge device based on the edge device's availability, communication protocol, maximum acceptable number of IoT devices and geographic location in the mobility model. If all the edge devices are disabled for any reason, the simulation will be stopped. And if certain edge devices are still running, but this IoT device cannot get connected with any running one, broker simply discards the request from IoT. Throughout the event processing, the connection header is transmitted with every event that sends or receives data which can be referred to as an information package. For example, when the edge datacenter broker wants to know where the EdgeLet come from, it will access the connection header to get the target id.

When the edge device is available to the IoT device, the broker will transmit the data to edge datacenter and then the edge datacenter will check again to ensure nothing is wrong. If anything goes wrong, the data will be discarded. In the case of all above process running smoothly, the edge device will finally process the EdgeLet, which is an asynchronous process. After processing the EdgeLet, the edge device will send an event to edge datacenter telling that the EdgeLet has been processed, and edge datacenter will send an event asking broker whether that IoT device is still available to this edge device or not. If the IoT device is not in the range, the broker will search the whole network to find an edge device that is connecting to the desired IoT device. This whole process is called edge-to-edge communication. After obtaining that edge device, the broker will indicate the edge device to transfer the returned data to another edge device i.e. having the connection with the desired IoT device. Presumably, this process does not cost any time in data transmission. Therefore, there is no delay, or battery consuming during this process. On the other hand, if everything goes fine, the processed data will finally be passed to the IoT device, then this IoT device will actuate, update battery, and check availability. After all this, the whole process is repeated. The simulation process is kept running until batteries run out, once all batteries ran out the simulation process will be terminated.

## 4 IoTSim-Edge Evaluation

To evaluate the applicability of IoTSim-Edge, three case studies are modeled. The details of each case study is given below.

### 4.1 Case 1: Healthcare System

Human activity recognition is beneficial for multiple purpose from maintaining fitness to monitoring healthcare . For example medical staff can easily monitors the health condition of a patient and offers the necessary assistance whenever required [18]. Based on the human physical activity, once can easily computes the calorie consumption and follows a particular diet or exercise [19]. Emergence of smart IoT devices such as *Pebble Watch* or *Fitbit* makes it more easier to monitor the activity. These smart devices creates raw activity data and sends to a smart phone using low energy protocols e.g. BLE (Bluetooth Low Energy). The raw activity data is processed by some algorithm to provide information about the activity levels.

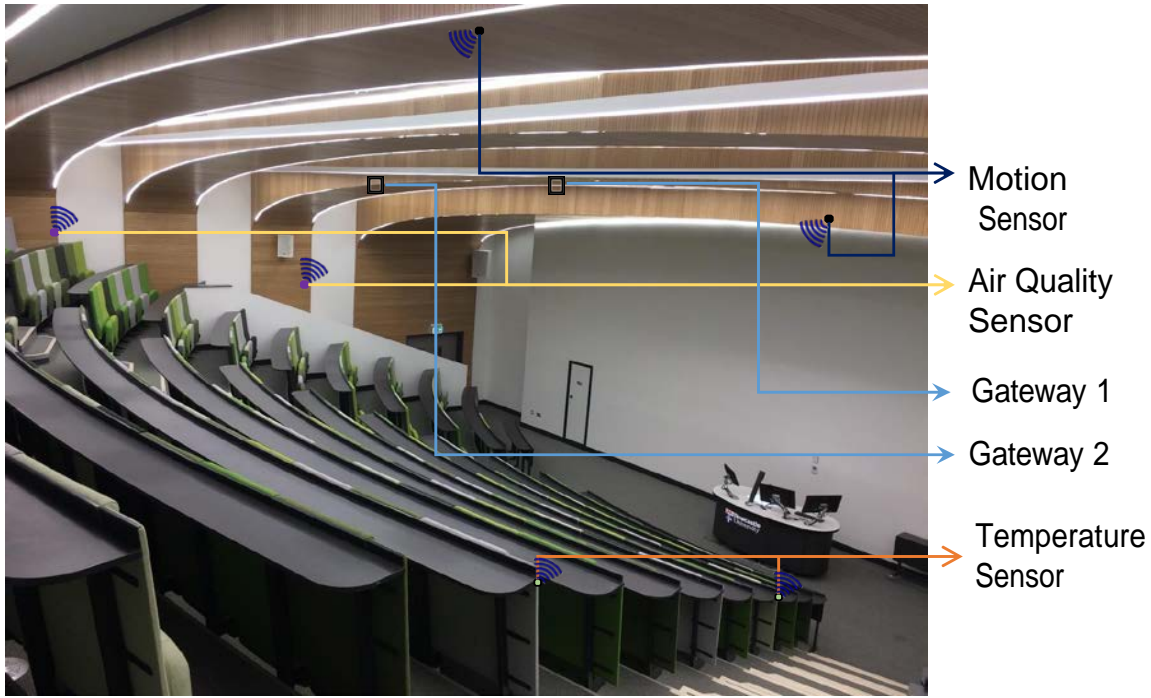

Fig. 12: Smart building example (Lecture hall of Urban Sciences Building at Newcastle University [1])

Following the work in [20] where the activity data is partitioned into multiple micro-operations, we also realized the same scenario. To process the raw data using step count algorithm, we model each micro-operation as a MEL that can be processed by underlying IoT or edge devices. Figure 10 shows the basic MEL graph for a simple healthcare scenario which in required to be deployed on the edge computing environment.

To find a pareto-optimal deployment solution on the distributed IoT infrastructure, it is necessary to analyze different possible deployment plans. Since there is strict dependency among MELs, it is important to consider this while deployment. To implement this in our simulator, we considered two edge devices in which one edge device has an embedded IoT device. The IoT device generates the data based on the defined data rate. Research shows that the battery drainage rate for transmission is higher than computation [20].

Since we considered edge devices are also battery powered, based on the processing happening on the edge device, the battery drainage rate varies which is analyzed in this work. However, performing more operations near to the IoT device will also increase the execution time which is very important in many cases. Our result also analyzes the total execution time in different cases. The configuration of experiment is shown in Table 3. Based on the real scenario [20], we assigned MIs to different MELs. Figure 11 gives the comparison of battery hours of edge to process the request with varying shrinking factor. Shrinking factor represents the processing happened on the edge device. The battery consumed by processing and communication is calculated as given in Equation 4.

The result in Figure 11 validates the proof that performing the processing on the edge device with IoT embedded will lead to save more battery as compared to sending all the raw data to other edge device. The result shows that processing 90% of operations on edge device increases the battery consumption by 266% as compared to performing only 10% of operation. Since we defined that edge E1 has small processing and storage capacity, all the processing cannot be performed there. Different type of analysis can be easily performed using this scenario. Users can also propose different algorithms to find a suitable deployment option based on multiple objectives in an easy way.

### 4.2  Case 2: Smart Building

Recently, smart building application that automates the lighting, heating, air-conditioning, air-quality, etc. has received a lot of attention [21]. Different type of sensors (e.g. temperature, humidity, motion, air quality) deployed at specific sites monitors the building activity  and sends the data to the associated edge device (e.g. raspberry pi or single personal computer) for processing and analysis. Figure 12 shows an example of smart building constructed at *Newcastle University*. Edge device do some local processing and sends the data to cloud if further storage or some complex analytics are required. Multiple IoT devices sends their data to one edge device following one specific communication protocol. Features like latency directly depends on the packet size and data rate. Research shows that one protocol is better than other protocol in terms  of packet size and data rate [22]. Again, since the IoT devices are battery operated, varying the data rate also changes the battery consumption. In the view of the fact that energy conservation is one of the important feature [23], recharging

TABLE 5: Configuration file for Case 3.

| IoT device | |
|---|---|
| Current location | 0, 0, 0 |
| IoT type | car |
| Movable | true |
| Data frequency | 1 |
| Data gen. time | 1 |
| Network protocol | Wi-Fi (802.11p) |
| IoT Protocol | XMPP |
| Max batt. capacity | 70   units |
| Battery drainage rate | 1   unit/hr |
| Number of entity | 'variable' |
| Velocity | 0.5 m/sec |

| Edge device 1 | |
|---|---|
| Type | Raspberry Pi |
| Location | 0, 0, 0 |
| Movable | False |
| Signal range | 50 m |
| Max IoT device capacity | 10000 |
| MIPS | 10000 |
| RAM | 10000 MB |
| Bandwidth | 10000 MBPS |

[Average latency for each response]

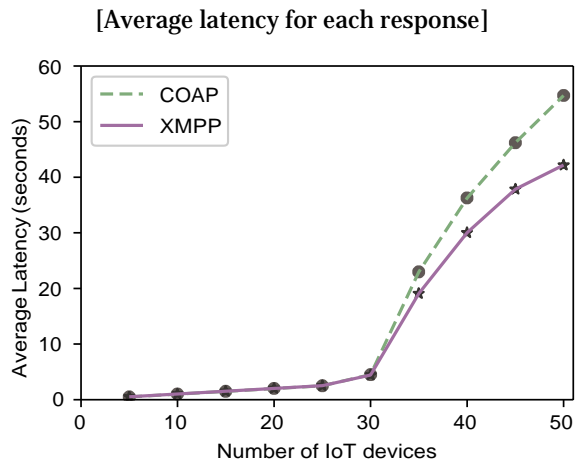

[Number of iteration before battery dies]

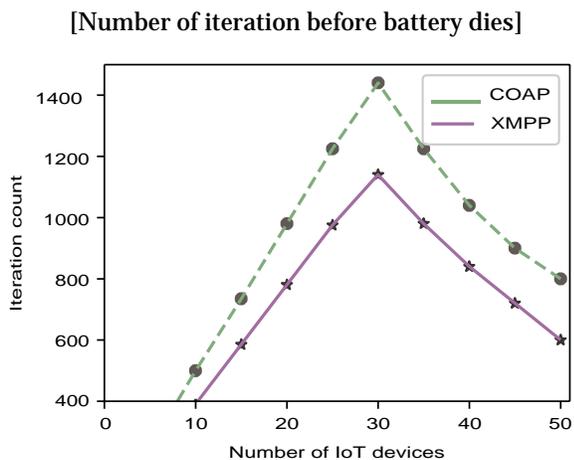

Fig. 13: Simulation result for smart building

or replacing the battery at a frequent rate is not suggested.

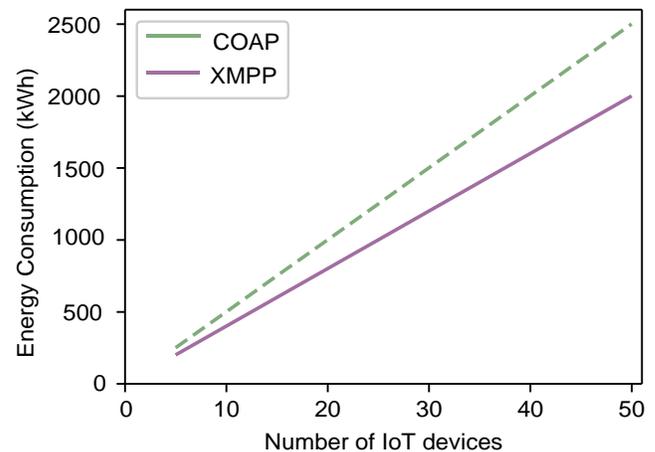

Fig. 14: Average energy consumption by each edge device

It is better to use a protocol that consume lesser energy.

To analyze this case, we have simulated a smart building scenario with varying number of sensors following either COAP or XMPP protocol. The configuration of IoT device is similar to the one discussed for Case 1 except the protocol for the executed IoT device is either COAP or XMPP and the IoT type is environmental as it monitors the building environment. Also, the number of IoT device varies from 1 to 50. The configuration of edge device for Case 2 is the same as explained in Case 1. The simulation results are presented in figure 13. Figure 13 shows the average latency for each response. The figure clearly shows that initially the latency increases slowly with increasing number of sensors however, after the edge device got saturated the latency becomes very high as the resource constrained edge device will take more time to process the request. The result also

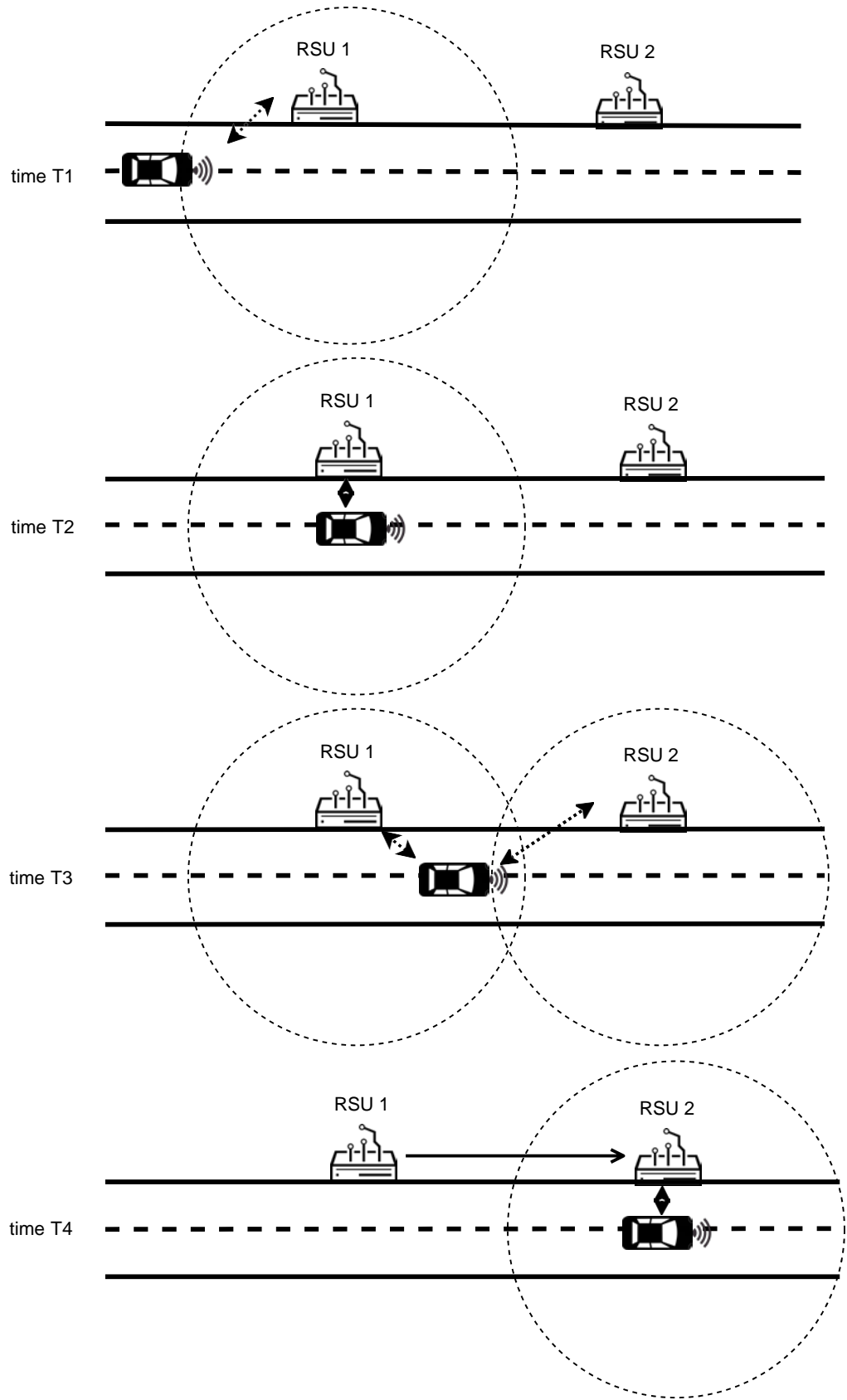

Fig. 15: Schematic diagram showing the car movement captured by the RSUs

shows that the latency of COAP increases at faster rate as compared to XMPP. The reason is justified from figure 13 where the iteration count (number of events generated during that particular scenario) for COAP is higher as compared

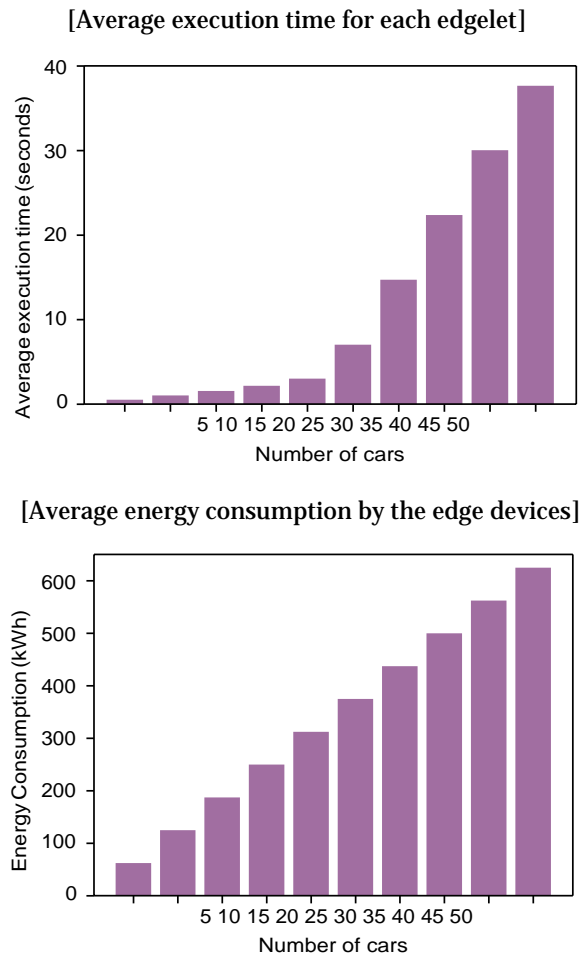

[Average execution time for each edgelet]

[Average energy consumption by the edge devices]

Fig. 16: Simulation result for RSUs

to XMPP. Since, COAP is a light weight protocol the battery also goes for longer time as compared to XMPP (see Table 4. The full battery drainage hour using both the protocols is given in Table 4.

Figure 14 shows the average energy consumption by the edge device for our test case. As we can see, the energy consumption of edge device increases with increasing number of IoT devices. However, the energy consumption for COAP is higher as compared to XMPP. This is because of the fact that using COAP more number of messages can be processed as shown in Figure 13. Also the energy always increases with increasing number of IoT devices regardless of the response time values.

### 4.3 Case 3: Capacity Planning for Road Side Units

Self driving cars [24] is an upcoming technology in which each car has an embedded sensor which communicates with the microcontrollers of RSUs to cooperatively maintain a smooth traffic flow and ensure the road safety. The RSUs can be smart edge devices that can process the data received from car and respond it back to the car to make some runtime decisions. The range of cars and RSUs are limited and a car is always moving with some velocity in a particular direction on the road. The coverage area of a RSU is dependent on the underlying transmission protocol. It is most probable at a point of time that the car losses its connection (handoff) with the previously connected RSU. The handoff may be hard or soft depending on the speed of car and distribution of the RSUs. Since the processed information from previous RSU is required to make certain decision, a RSU to RSU (edge-to-edge) communication is also established. Based on the direction of the car movement, one RSU sends data to another RSU which is further delivered to the specified car. Since the range and processing capacity of RSU is limited, it is necessary to analyze how many car data can it process without losing any information [25]. Also, it is necessary to guarantee the QoS requirement e.g. response time of the application.

We modeled this scenario using our simulator. Figure 15 shows the schematic diagram of the model considered by our simulator. At time *t1*, the car found a new RSU. It first establishes a connection with the new RSU, *RSU 1* and then start sending data. At time *t3*, it reaches at a position where it is in the range of two RSUs. Based on it's movement, it start making connection with *RSU 2* while still sending data to *RSU 1*. When it lost connection with *RSU 1*, it start sending data to *RSU 2*. *RSU 1* figures out that the car is now not in their range and based on the velocity information, it transmits the processed data to *RSU 2* which delivers the information to the car as shown for time *t4*. This scenario validates the mobility feature as well as cooperative edge communication feature of our simulator. Table 5 gives the specific simulation configuration of the specified case. Both the edge devices are identical except the location ($E1$ is at $0,0,0$ while $E2$ is at $50,0,0$).

The simulation result is presented in Figure 16. The result in Figure 16 shows the average execution time with respect to the number of cars. The execution time increases as the number of car increases. With the increase in the number of cars, the number of connection also increases. Since the number of edge devices are constant and the processing is done in a time-shared manner, requests are queued before processing which leads to the increased execution time. Also with the mobility of car, if the car move away from the range of one RSU, then that RSU has to send the processed data to other RSU which further sends to the car. Since this is also added to the actual processing time, the execution time is higher. To process increased number of requests from car, edge will consume more energy as verified by our simulation (see Figure 16) which shows that the average energy consumption of edge device increases with increasing number of cars (IoT).

## 5 Related Works

With the surge in interest in IoT and edge computing, numerous simulation tools are developed in the past few years. Some of the tools are extended from the existing network and cloud simulators, however, there is a gap in the existing simulators and the realistic modeling of edge and IoT environment. In this section, first, we discuss the currently available simulation tools for the network, cloud, and IoT environment and how they are not able to model the existing IoT-Edge environment. Further, we show how our simulation framework is able to satisfy the available challenges in a holistic way.

Network Simulators Several simulation tools have been proposed for simulating computer network in the last decade. Among them, here we discuss a few well-known network simulation tools. OMNeT++ [26] is a C++ based discrete event simulation framework that can simulate a distributed processing environment with network communications. It supports parallel simulation as well as real-life protocol implementation in the simulation models. It also supports the simulation of some of the network protocols but does not support edge communication protocols. Castalia [27] is developed on top of OMNet++ platform to simulate Body Area Network and other related networks which consists of lower-energy powered devices. It also endures in simulating a large number of mobile nodes dynamically. By extending this Benedetti et al. [28] proposed GreenCastalia to support energy-harvesting techniques in the simulator. However, this does not support the modeling of different communication protocols. TOSSIM [29] is an open source even-driven simulator to simulate wireless sensor networks (WSN). It uses the NesC programming language to develop the simulation environment. TOSSIM supports features such as sophisticated network connectivity, bridging, and scalability. However, this simulator does not support energy consumption and mobility modeling. To support the modeling of power consumption in WSN another simulator was proposed which is known as Power-TOSSIM z [30] by extending TOSSIM. PowerTOSSIMz incorporates the non-linear behaviour of the battery model simulation. However, it does not support the mobility feature. NS-3 [31] is another C++ based discrete event simulation tool which has a Python scripting interface. This tool can simulate a distributed environment with virtualization support. But NS3 is not suitable for IoT simulation at edge level since it does not support the scheduling and application composition features.

Cloud Simulators Many simulation tools have proposed so far for the simulation of the cloud computing environment. Among them, CloudSim [7] is the most used cloud simulator among the research community. CloudSim is event-driven simulation tools which support both behaviour and system modeling of cloud environment components. However, CloudSim does not support simulation and modeling of IoT and edge environment. The proposed IoTSim-Edge is developed on top of CloudSim simulator by extending it.

iCanCloud [32] is another cloud simulator which supports large scale simulation with support of communication and physical models. This simulator has customizable global hypervisor which can support any cloud brokering policy. It contains Amazon public cloud instances and supports MPI. NetworkCloudSim [33] is another simulator that allows us to model the network behaviour of cloud datacenters. However, these simulators do not support IoT and edge simulation.

Another cloud simulator is GreenCloud [8] which is extended from the NS2 simulation tool. GreenCloud is a packet level simulation tool which can measure the energy consumption of datacenter components. This simulator only focuses on the calculation of energy consumption to ensure energy-aware placement. It also does not support IoT and edge simulation. DCSim [34] is another cloud simulator which is specifically focused on IaaS cloud environment simulation. It supports modeling and simulation of the datacenter, host and VM with a limited number of application and resource management policies. However, simulation of IoT and edge environment is not supported by this simulator.

IoT and edge Simulators Numerous simulation tools have been proposed for simulating either IoT or edge environment. SimIoT [35] is a simulation tool which models the communication between IoT devices and cloud datacenter but they did not incorporate edge devices in the simulation. This simulator enabled experimentation of multi-user submission dynamically in the IoT environment. However, this tool did not consider the heterogeneity of IoT devices. Also, the energy consumption of IoT devices is ignored here. Österlind et al. proposed a cross-level sensor network simulator called COOJA [36] which can be used to simulate different deployment levels (machine code instruction set level, operating system level, and network level). This tool is suitable for heterogeneous network nodes simulation but is designed only for the Contiki operating system.

iFogSim [10] support modeling of Fog and IoT environments and measure the impact of resource management techniques in network congestion, latency, cost, and energy consumption. This simulation tool considers edge devices as Fog devices. However, they did not consider edge communication protocols in their simulation tool. The IoTSim [37] support simulation of Map-Reduce process which is known as a big data processing paradigm. This simulation tool is only used for modeling Map reduce processes. EdgeCloudsim [9] is another edge simulation tool build on the top of CloudSim toolkit. It supports computational and networking aspects of edge computing paradigm. EdgeCloudSim provides network link model, the mobility model, and edge server model to evaluate the aspects of edge computing. However, it is not bale to model the application composition of IoT application. DISSECT-CF [38] is another simulation tool built on top of the cloud computing concept, which can evaluate the energy consumption of IaaS (Infrastructure as a service), model scheduling and internal infrastructure behaviours [38]. However, Kecskemeti, et al. [6] stated that "Although the integrated sensor models are generic, they might still not be inapplicable in future IoT scenarios".

In summary, all above-discussed simulators do not support edge communication protocols and energy calculation (battery power). Also most of them are not able to define the application composition in IoT environment. These features are essential for any IoT application. A composite simulation environment for IoT-Edge is crucial to help researchers and industries to gain the real potential of edge processing. Our proposed simulator, IoTSim-Edge covers all these features in a holistic manner. The advantage of our IoTSim-Edge as compared to the existing simulators is clearly visible from Table 6 as the proposed simulator is able to satisfy more features.

# 6 CONCLUSION AND FUTURE WORKS

The recent advances in IoT and edge computing offers numerous services for the users near to the edge. To enable these services, we need to develop new methods and techniques which is required to be tested before offering to the users. The deployment of these methods and techniques for IoT application in the real environment is a complex, time-consuming task and also not cost effective. Moreover, the application requires massive data collection and processing at the autonomous end devices which need proper validation before deploying in the real environment. We also need to test the scalability and usability of the new methods and techniques.

TABLE 6: Comparison of various open-source simulation tools that are currently using for IoT or edge simulation.

| Simulator Name | Features | | | | | | | |
|---|---|---|---|---|---|---|---|---|
| | Network Communication | Heterogeneity | Edge communication protocols | Network Protocols | Mobility | Battery Power | IoT Devices | Edge Processing |
| OMNeT++ [26] | ✓ | ✓ | | ✓ | | | | |
| TOSSIM [29] | ✓ | ✓ | | ✓ | | | | |
| PowerTOSSIMz[30] | ✓ | ✓ | | ✓ | | ✓ | | |
| CloudSim [7] | ✓ | ✓ | | | | | | |
| iCanCloud[32] | ✓ | ✓ | | | | | | |
| GreenCloud [8] | ✓ | ✓ | | | | | | |
| DCSim [34] | ✓ | ✓ | | | | | | |
| NS-3 [31] | ✓ | ✓ | | ✓ | | | | |
| SimIoT [35] | ✓ | | | | | | ✓ | |
| Cooja [36] | ✓ | ✓ | | | | | ✓ | |
| iFogSim [10] | ✓ | | | | | | ✓ | ✓ |
| IoTSim [37] | | | | | | | ✓ | |
| EdgeCloudSim [9] | ✓ | | | | ✓ | | | ✓ |
| DISSECT-CF [38] | | | | | | | ✓ | |
| IoTSim-Edge (Proposed) | ✓ | ✓ | ✓ | ✓ | ✓ | ✓ | ✓ | ✓ |

Because of the efforts of the researchers in distributed computing domain, there are many simulators currently available for network, cloud and IoT simulation. However, these simulators are not perfectly suitable for IoT and edge due to some exclusive features of edge computing. These features are including highly heterogeneous devices, communication protocols and mobility. To incorporate all these exclusive characteristics, we proposed a novel IoTSim-Edge simulator which models numerous features including (i) device heterogeneity, (ii) application composition, (iii) variety of IoT communication protocols, (iv) device movement and mobility and (v) battery features. It benefits researchers to develop their own prototype and test in a scalable simulation environment. They can identify the bottlenecks and test the performance of their methods and techniques with no cost which will help them to improve the usability and performance of their proposed techniques.

We validated the effectiveness of the simulator by considering three test cases: healthcare system, smart building, and capacity planning for roadside units. The result showed the varying capability of IoTSim-Edge in terms of application composition, battery-oriented modeling, heterogeneous protocols modeling and mobility modeling along with the resources provisioning for IoT application in an efficient manner.

As the future works, we will develop an IoT-Edge emulator which consists of all the above features and integrate with the IoTSim-Edge. This can help to test all these functionality in a realistic environment.